\documentclass{article}
\usepackage{epsfig,multicol}
\usepackage{fleqn,espcrc2}
\title{\bf WIGNER LATTICE ORDER, COLLECTIVE MODE, AND SUPERCONDUCTIVITY IN LA$_{1.985}$SR$_{0.015}$CUO$_{4+\delta}$ SYSTEM}

\author{YOUNG HOON KIM$^{1}$ and PEI HERNG HOR$^{2}$\\ 
$^{1}$ Department of Physics, University of Cincinnati, Cincinnati, Ohio 45221-0011, U.S.A.\\
$^{2}$ Department of Physics and Texas Center for Superconductivity,
University of Houston\\ Houston, Texas 77204-5002, U.S.A.\\
\vspace{12pt}
\it Published on Modern Physics Letters B, Vol. 15, No. 15 (2001) 497-513 }

\begin{document}

\begin{abstract}
We have studied far-infrared charge dynamics of Sr- and O- co-doped La$_{1.985}$Sr$_{0.015}$CuO$_{4+\delta}$ with $\delta$ = 0.024 ({\it p} = 0.063 per Cu) and $\delta$ = 0.032 ({\it p} = 0.07). We found that two-dimensional Wigner lattice order is the ground state of cuprates away from half-filling. We found that the presence of 2D Wigner lattice and the pinned Goldstone mode is essential for the cuprate physics and superconductivity. We propose that all the high T$_c$ physics are based on the existence of these peculiar 2D electron lattices.
\end{abstract}

\maketitle

Superconductivity in layered copper-oxides (cuprates) has been
one of the important problems in condensed matter physics that
still awaits the answer for the underlying mechanism(s) that
drives this doped insulator to a superconductor with high
transition temperature (T$_{c}$). Theoretically, the main problem
has been finding an adiabatic connection between the
antiferromagnetic order and superconducting order in the quasi
two-dimensional (2D) CuO$_{2}$ planes upon hole doping. On the
antiferromagnetic insulating side, the t-J model,$^{1}$ which
is derived from the Hubbard model,$^{2}$ has been believed to
capture the essential physics of CuO$_2$ planes. As the system is
doped with holes, Anderson proposed$^{3}$ that the
Mott-Hubbard insulator becomes a 2D Luttinger liquid where the
spin and charge of electrons are separated into spinons and
holons. Later, a philosophically different SO(5) theory was
proposed by Zhang,$^{4}$ which is a microscopic model with a
five-component order parameter that embraces both the
antiferromagnetic order corresponding to a charge zero, spin 1
particle-hole pair at half-filling ({\it p} = 0) and the
superconducting order corresponding to a charge $\pm$ 2e cooper
pair spin singlet away from half-filling. Experimentally, besides
antiferromagnetism and superconductivity, spin clusters, spin
glass and a highly elusive hole concentration
({\it p})-independent pseudogap in spin excitation and charge
spectra are observed in the normal state.$^{5}$ Based on the
observed rich physical phenomena, a generic electronic phase
diagram has been generated as a function of temperature and hole
concentration.$^{6}$ Although it is widely recognized that
studying this phase diagram holds the key to understand the
cuprate physics, until now no theory has been able to account for
the mere existence of such a phase diagram.

In a recent cation and anion co-doping study of the general
properties of the electronic phase diagram, incipient electronic
instabilities related to energetically favored electronic phases
were identified at special hole concentrations ({\it p}$_s$).$^{7}$ In particular, a {\it p}$_s$ = 1/16 is observed. This
{\it p}$_s$ = 1/16 corresponds to a previously reported critical
hole concentration ({\it p}$_c$) which dictates the doping
efficiency and signals a chemical potential jump of the doped
holes.$^{8}$ This {\it p}$_c$ = 1/16 is also the point where
superconductivity first appears in the phase diagram. In Fig.1 we
present the onset T$_c$ vs. {\it p} curves of
La$_{2-x}$Sr$_x$CuO$_{4+\delta}$ for x = 0, 0.015, and 0.05. The
T$_c$ is defined as the onset temperature of the Meissner
effect. The {\it p} of each sample series is varied precisely
by small steps of $\Delta${\it p} = 0.0025 $\sim$ 0.004 using
electrochemical oxidation to change the oxygen concentration,
$\delta$.$^{8}$ While a single 30K transition is observed for
x = 0 samples, a dramatic increase of T$_c$ from $\sim$ 15K to $\sim$
30K is clearly seen for both x = 0.015 and 0.05 samples. In
particular, for the x = 0.015 sample it is remarkable to observe
an almost discontinuous jump from T$_c$ $\sim$ 16K to 26K by a
very small change of carrier concentration $\Delta${\it p} =
0.004 near {\it p}$_c$. Since {\it p}$_c$ = 1/16 is a special
carrier concentration related to a possible electronic phase
instability, far-infrared studies of the x = 0.015 sample at
various {\it p} across {\it p}$_c$ provides a unique
opportunity to investigate the specific role of this instability
to the normal state charge dynamics and to the occurrence of the
superconductivity in cuprates.

\begin{figure}[tbh]  
\begin{center}
\epsfig{file=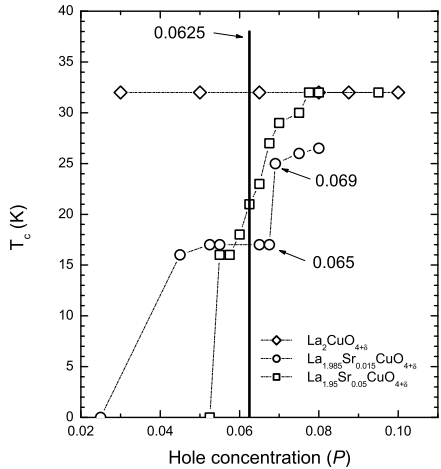,bb=11 14 221 236,width=5.8cm,clip}
\caption{Superconducting onset temperature (T$_c$) versus hole
concentration ({\it p})} 
\end{center} 
\end{figure}

The nature of charge carriers in the CuO$_2$ planes
(ab-planes) and their interactions may be directly probed by
far-infrared reflectivity measurements. However, because of the
Drude-like free-carrier contributions to the far-infrared
conductivity, the normal state reflectivity is rather
featureless. The only persistent characteristic spectral feature
found was the emergence of a doping-independent "knee"-like
structure in the reflectivity at $\sim$ 400 cm$^{-1}$ ($\sim$ 50
meV) upon cooling the sample below T$_c$.$^{9}$  This structure
manifests itself as a broad dip in the frequency dependent
conductivity for T $<$ T$_c$. The physical origin of the dip has
not been resolved.$^{10,11}$ Recently, the dip in
YBa$_{2}$Cu$_{3}$O$_{7-\delta}$ systems has been attributed to the
coupling of electrons to the 41 meV resonance peak observed in
neutron scattering.$^{12}$ Also, it was found that there
exists substantial residual conductivity in the far-infrared
below the dip frequency ($\omega$) $\sim$ 400 cm$^{-1}$ even for T
$<$ T$_c$. The physical nature of this residual background
conductivity and its relevance to the superconductivity has not
been understood.

Studies of photoinduced carriers in undoped cuprate provide an
alternative approach to understand the nature of charge carriers
and their interactions with the lattice. Because there is no
Drude conductivity, one can clearly discern the carrier-induced
background conductivities. Although the spectral features for
$\omega$ $<$ 200 cm$^{-1}$ could not been resolved due to the
nature of the experiments, the infrared photoinduced absorption
measurements of La$_2$CuO$_4$ and Nd$_2$CuO$_4$ revealed that the
photocarriers became self-localized via coupling to the lattice
as evidenced by the photoinduced infrared-active vibrational
modes.$^{13}$ Also, there existed an accompanying electronic
excitation peak at $\sim$ 1000 cm$^{-1}$ ($\sim$ 0.12 eV) with an
onset at $\sim$ 400 cm$^{-1}$. This result was interpreted as the
result of phase separation of the photocarriers in the CuO$_2$
planes into hole-rich domains with local {\it p} comparable to that of
optimally doped cuprates.$^{14}$

In this work, we investigated two polycrystalline x = 0.015
samples at either side of the T$_c$ jump near {\it p}$_c$. In
preparing the sample, different electrochemical oxidation
techniques have been carefully evaluated$^{15}$ and the
thermodynamic equilibrium properties of the electrochemical
intercalation of oxygen into polycrystalline
La$_{2-x}$Sr$_{x}$CuO$_{4+\delta}$ system have been studied.$^{16}$ Because of the strong anisotropic nature of cuprates
where the electromagnetic response is nearly insulating for
polarization parallel to the c-axis and conducting for
polarization parallel to the ab-plane and because the c-axis
far-infrared properties of single crystalline
La$_{2-x}$Sr$_x$CuO$_4$ are well-documented,$^{17}$ we were
able to unambiguously identify the in-plane far-infrared charge
dynamics of polycrystalline
La$_{1.985}$Sr$_{0.015}$CuO$_{4+\delta}$ sample, which is
directly relevant to the normal and superconducting states.

We have carried out far-infrared reflectivity measurements of
two different samples of La$_{1.985}$Sr$_{0.015}$ CuO$_{4+\delta}$
near {\it p}$_c$ as a function of temperature for $\omega$
between $\sim$ 7 cm$^{-1}$ and 5000 cm$^{-1}$; one with $\delta$ =
0.024 (or {\it p} = 0.063 per Cu) which undergoes the
superconducting transition at T$_c$ = 16 K and the other at
$\delta$ = 0.032 (or {\it p} = 0.07 per Cu) with T$_c$ = 26 K
defined as the onset temperature of the Meissner effects shown in
the inset of Figure 2a and 2b, respectively. Both resistivity
$\rho$ vs. T and d$\rho$ /dT vs. T curves are depicted in Figures
2a and 2b. Although they have quite different T$_c$'s, one
interesting feature in both d$\rho$ /dT vs. T curves is a common
minimum and maximum observed at $\sim$ 200K and $\sim$ 150K.
These very similar behaviors suggest that the same physics is in
operation for the normal state in both samples. We noted that the
Meissner effect signal size of $\delta$ = 0.032 sample is larger
than that of $\delta$ = 0.024 sample. This result indicates that
the superconducting state in the $\delta$ = 0.032 sample is not
due to the formation of oxygen-rich phase. A smaller Meissner
effect is expected since there will be a smaller superconducting
volume for samples chemically phase-separated into oxygen rich
domains. Indeed, there are only two T$_c$'s, one at $\sim$ 15 K
and the other at $\sim$ 30 K, observed for the entire underdoped
regime in the pure oxygen doped equilibrium samples.$^{18}$
This indicates, besides two chemical phases, that two T$_c$'s
might be coming from two different electronic phases.

\begin{figure}[tbh]  
\begin{center}
\epsfig{file=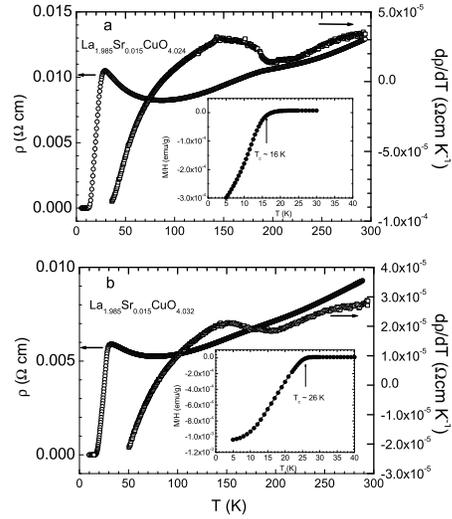,bb=14 12 235 272,width=5.8cm,clip}
\caption{Resistivity and first derivative of resistivity versus
temperature curves for oxygen co-doped polycrystalline} 
\end{center} 
\end{figure}

In this experiment, we directly measured the sample temperature
from the backside of the sample. The temperature resolution was
(1 $\pm$ 0.1) K for T $<$ 40 K. Far-infrared properties,
represented by a complex dielectric function 
\begin{equation}
$$\epsilon(\omega)
= \epsilon_1(\omega) +
4\pi i \sigma_1(\omega)/\omega$$
\end{equation}
have been calculated from
a Kramers-Kronig analysis of the measured reflectivities. For
$\omega$ $<$ 7 cm$^{-1}$, we used an empirical extrapolation
scheme R($\omega$) $\approx$ 1 - C$\omega^{\frac{1}{2}+\delta}$
with -1/2 $\leq$ $\delta$ $\leq$ 1/2 subject to
$\sigma_{1}$($\omega$ $\rightarrow$ 0) = $\sigma_{dc}$ where
$\sigma_{dc}$ is the measured dc conductivity.$^{19}$

Far-infrared reflectivities of $\delta$ = 0.024 and $\delta$ =
0.032 samples are shown in Figure 3 at selected temperatures.
Besides the contribution due to the well-known in-plane mode at
$\sim$ 500 cm$^{-1}$ and the intense c-axis phonon mode at
$\omega$ $\sim$ 220 cm$^{-1}$ and a weak mode at $\sim$ 350 cm$^{-1}$,
there appears a sharp minimum in the reflectivity at $\omega$
$\sim$ 16 cm$^{-1}$ that resembles the reflectivity curve in the
vicinity of the plasma edge of a free electron gas and several
unusual notable features. Below the minimum at $\omega$ $\sim$ 16
cm$^{-1}$ the reflectivity approaches unity as $\omega$ approaches zero.
For $\delta$ = 0.024 sample, a bump at $\omega$ $\sim$ 25
cm$^{-1}$ is evident and another weak broad feature located at
$\omega$ $\sim$ 100 cm$^{-1}$ on which a sharp peak develops as T
decreases. We noted that the reflectivity does not approach unity
for frequencies down to $\omega$ $\sim$ 7 cm$^{-1}$ even in the
superconducting state.

\begin{figure}[tbh]  
\begin{center}
\epsfig{file=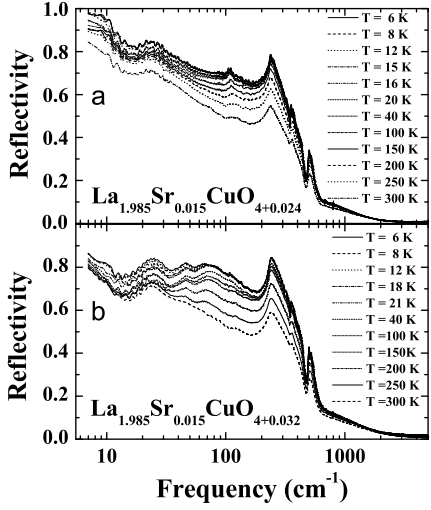,bb=11 12 225 263,width=5.8cm,clip}
\caption{Reflectivity versus frequency curves at various
temperatures: (a) $\delta$ = 0.024 sample and (b) $\delta$ =
0.032 sample.} 
\end{center} 
\end{figure}

For $\delta$ = 0.032 sample, in addition to what have already
been seen for $\delta$ = 0.024 sample, two additional bumps are
present: one at $\omega$ $\sim$ 45 cm$^{-1}$ and the other at
$\omega$ $\sim$ 80 cm$^{-1}$. One remark on the reflectivity
data is that because we use polycrystalline sample, there exist
possible errors in determining the absolute reflectivity at
$\omega$ $\rightarrow$ 0 due to the overestimation of the
reflectivity. We were unable to determine the reflectivity of
$\delta$ = 0.032 sample for $\omega$ $<$ 10 cm$^{-1}$ to the
desired accuracy ($<$ 1 \%). During the course of this
experiment, we found that the reflectivity in the spectral range
between 100 cm$^{-1}$ and 600 cm$^{-1}$ are close to true
reflectivity of the sample. The uncertainty of the far-infrared
reflectivity of $\delta$ = 0.032 sample for $\omega$ $<$ 10
cm$^{-1}$ is slightly higher than that of $\delta$ = 0.024 sample
($\sim$ 1$\%$). However, this error does not affect our
conclusions for $\omega$ $>$ 10 cm$^{-1}$.

The Kramers-Kronig derived real part of conductivity
$\sigma_{1}$($\omega$) and the real part of dielectric function
$\epsilon_{1}$($\omega$) of $\delta$ = 0.024 sample are displayed
in Figure 4a \& 4b at various T, respectively. In
$\sigma_{1}$($\omega$), there appear a sharp upturn approaching
$\sigma_{dc}$ in $\sigma_{1}$($\omega$) below 10 cm$^{-1}$ and an
extremely sharp (FWHM $\sim$ 10 cm$^{-1}$) peak located at
$\omega$ $\sim$ 23 cm$^{-1}$ whose strength increases as T
decreases. We can also observe a broad mode at $\sim$ 90
cm$^{-1}$ that becomes pronounced at low T and an electronic
excitation peak at $\sim$ 1000 cm$^{-1}$ with an onset at $\sim$
400 cm$^{-1}$. All these features exist in addition to the
well-known ab-plane and c-axis CuO$_2$ phonon modes. We identify
this electronic excitation peak with the lower energy electronic
excitation in the photoinduced infrared absorption measurements.
All these features in $\sigma_{1}$($\omega$) develop on top of a
constant, rather frequency independent background conductivity of
$\sim$ 10 $\Omega^{-1}$cm$^{-1}$ which presumably comes from the
c-axis contribution.

\begin{figure}[tbh]  
\begin{center}
\epsfig{file=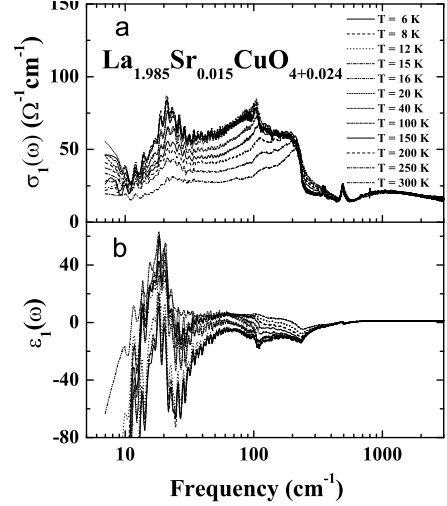,bb=11 11 222 256,width=5.8cm,clip}
\caption{Kramers-Kronig derived (a) real part of the conductivity
$\sigma_{1}$($\omega$) and (b) real part of the dielectric
function $\epsilon_{1}$($\omega$) for $\delta$ = 0.024 sample.} 
\end{center} 
\end{figure}

From the $\epsilon_{1}$($\omega$) plot in Figure 4b, it is
clear that the reflectivity minimum at $\sim$ 16 cm$^{-1}$ arises
from the zero crossing of the $\epsilon_{1}$($\omega$) and there
is no sign of any corresponding structure seen in the
conductivity. Therefore, we assign the reflectivity minimum as
the plasma edge due to the free carriers with screened plasma
frequency $\omega_{p}$ $\sim$ 16 cm$^{-1}$. By assuming the free
carrier mass to be the electron mass and using the static
dielectric constant $\epsilon_{1}$(0) $\sim$ 40 estimated from
$\epsilon_{1}$($\omega$) at 300 K, we find the free carrier
concentration n$_F$ $\sim$ 2.3 x 10$^{17}$ carriers$/$cm$^3$
which is only $\sim$ 0.4 \% of the total carriers in the system.
Even though the measured $\sigma_{dc}$ is in the range of $\sim$
100 $\Omega^{-1}$cm$^{-1}$ and the observed $\sigma_{1}$($\omega$)
is the average between the ab-plane and c-axis conductivities, the
sharp upturn in $\sigma_{1}$($\omega$) toward $\sigma_{dc}$ at
$\omega$ $<$ 10 cm$^{-1}$ indicates that the free carriers
experience virtually no scattering ($\Gamma$ $\sim$ 4 cm$^{-1}$).
This observation suggests that the extremely small fraction of
free carriers participate in a coherent charge transport in
La$_{1.985}$Sr$_{0.015}$CuO$_{4+\delta}$.

The $\epsilon_{1}$($\omega$) of $\delta$ = 0.024 sample clearly
reveals that the $\omega$ $\sim$ 23 cm$^{-1}$ mode is a collective
mode ($\omega_{GL}$) and the 400 cm$^{-1}$ is the corresponding
single particle excitation gap (2$\Delta$). One might consider
that the origin of this collective mode is from a spin density
wave (SDW). However, in this event the entire spectral weight is
expected to move to the collective mode of the SDW because the
dynamic mass of the SDW is the free electron mass. That is, we do
not expect to have the single particle excitation peak.
Therefore, this collective mode is resulting from the charge
condensation into a one-dimensional or 2D density wave in the
CuO$_2$ planes and this density wave is pinned as evidenced by
the finite frequency of the collective mode.

The spectral weight distribution indicates that almost all the
carriers introduced to the CuO$_2$ planes are spent to form a
pinned charge density wave (CDW). Since a pinned CDW is an
insulator,$^{20}$ the pinned CDW itself cannot give rise to
the observed nearly dissipationless metallic conductivity.
Therefore, the coexistence of a pinned CDW and metallic
conductivity demands a channel for coherent charge transport. In
order to defy all the scattering with the CuO$_2$ phonons, the
free carriers must "ride" the density wave. In the case of a
one-dimensional CDW, formation of topological charged solitons
(phase kinks) is a possibility.$^{21}$ However, because of
the topological nature of the solitons, it requires that both
soliton and antisoliton must hop to the neighboring CDW chains to
transport charges. Therefore, the observed coherent charge
transport will not take place through the soliton hopping
mechanism. In the stripe model,$^{22}$ each stripe in the
CuO$_2$ plane is considered as a river of charge and conducting.$^{23}$ Hence, no charge gap or pinned collective modes are
expected. This point has been verified through the transport
$^{24}$ and infrared studies of static charge stripes in
La$_{1.6-x}$Nd$_{0.4}$Sr$_x$CuO$_4$ systems.$^{25}$ Therefore,
our observation bears neither the signature of the idea of
strings of charges, either static or dynamic, nor the
one-dimensional CDW's in the CuO$_2$ plane.

The other possibility is a 2D Wigner lattice.$^{26}$ The
observed data may imply that charges (holes in our case) added to
the CuO$_2$ planes condense to form a 2D Wigner lattice. The
repulsive Coulomb interaction between the holes at dilute
concentration enables the holes to form a 2D square lattice in
the CuO$_2$ plane. In this model, the $\omega$ = 0 Goldstone mode
of the Wigner lattice has been shifted to $\omega_{GL}$ $\sim$ 23
cm$^{-1}$ due to the commensuration pinning of the Wigner lattice
to the underlying CuO$_2$ lattice. Charge-lattice coupling would
enhance the lock-in of the Wigner lattice with the underlying
CuO$_2$ lattice$^{13}$ and stabilize the square lattice by
lowering the total energy.

Based on the study of the electronic phase diagram at well
above room temperature (85 $^{o}$C) where an intrinsic electronic
phase instability is observed at exactly {\it p}$_c$ = 1/16. We
can now formally relate our low temperature results to the high
temperature special carrier concentration. It seems that this
instability survives to low temperature and we suggest that all
the doped holes in the CuO$_2$ plane are condensed to form a
(4x4) square Wigner lattice with sides L = 4a (a = Cu-Cu
distance). Then, the observed sharp Drude-like free-carrier
conductivity in this insulating Wigner lattice ground state comes
from the coherent charge transport of the excess holes
$\Delta${\it p} = {\it p} - {\it p}$_{c}$ $\sim$ 0.0005 per
copper which corresponds to $\sim$ 0.8\% of the total carriers in
$\delta$ = 0.024 sample. In order to achieve such a coherent
transport of charges, the free-carriers must reside in a state
where the scattering with the phonons of the CuO$_2$ lattice can
be avoided. That is, the free carriers need to "ride" the Wigner
lattice. In a square lattice model, when an excess hole is
introduced to the Wigner lattice, it will find an energetically
lowest location on the lattice. Because of the on-site Coulomb
repulsion at the lattice sites, the minimum energy location would
be the center of the square lattice. At this location, the
excess hole seats in a local harmonic potential with a zero point
energy. This hole now may move because it "sees" a periodic
harmonic potential in the directions of the neighboring identical
energy minima. This causes the zero point energy of the excess
holes to be broadened into a band with the width of which is
density dependent. This band is gapped by the Coulomb interaction
energy at the minimum energy site. For charge carriers in the
Coulomb band, the main energy dissipation channel for electrons
is through the scattering with acoustic phonons of the Wigner
lattice. At
high temperatures, there exists possible promotion of holes from
the Wigner lattice site to an interstitial position and vice
versa. A similar idea of an interstitial band has been discussed
by de Wette for a three-dimensional electron Wigner lattice.$^{27}$

In this band picture, we anticipate two absorption peaks in
addition to the transition across the single particle excitation
gap: One due to the transition ($\omega_1$) from the Coulomb band
to the single particle excited state and the other involves the
transition from the ground state to the Coulomb band ($\omega_2$).
However, the $\omega_2$ transition from the ground state to the
Coulomb band requires an energy across the gap less the Coulomb
energy change in the lattice which is essentially the same as the
static Coulomb energy difference between the occupied lattice
site and the interstitial lattice site. We suggest that
$\omega_2$ is of the order of $\sim$ 0.012 eV ($\sim$ 100
cm$^{-1}$) as seen in the oxygen gas effusion experiments.$^{28}$ This implies that the Coulomb energy difference
between the occupied lattice site and the interstitial lattice
site is $\sim$ 0.025 eV. Therefore, we identify the 90 cm$^{-1}$
peak with $\omega_1$ with a full width W$_c$ $\sim$ 200 cm$^{-1}$
(0.025 eV) and the Coulomb band gap is $\sim$ 300 cm$^{-1}$. Note
that $\omega_1$ and $\omega_2$ are quite close.

The effective mass of the Wigner lattice (m$^*$) can be
estimated by taking the ratio of the oscillator strength of the
collective mode (S$_{GL}$) to that of the single particle
excitation (S$_{2\Delta}$) calculated from the
$\sigma_{1}$($\omega$) and by applying the sum rule
\begin{equation}
$$S_{GL} + S_{2\Delta} + S_{other} = \pi ne^2/2m_e$$
\end{equation} 
with
\begin{equation}
$$S_{GL} = \int\sigma_{1}^{GL}(\omega)d\omega = \pi n_{W}e^2/2m^*$$
\end{equation}
and S$_{other}$ is the oscillator strength of all other
excitations. Using the experimentally estimated oscillator
strengths, we found m$^*$ $\sim$ 170 m$_e$ at 300 K, $\sim$ 80
m$_e$ at 150 K and $\sim$ 60 m$_e$ at 30 K. This implies that the
density waves develop the long-range order at T $>$ 150 K by
drawing its strength from the single particle excitations. We
shall see later in the oscillator strength vs. T plot that the
crossover occurs at T$_0$ $\sim$ 200 K. However, we believe that m*
is underestimated from the experimental data because the measured
reflectivity for frequencies above 400 cm$^{-1}$ is less than
true reflectivity due to the ceramic nature of the sample. In
addition, symmetric band is assumed to take into account of the
spectral weight above 0.5 eV. Fitting for each peak was made
using a symmetric Gaussian function as an approximation without
taking into account of asymmetry.

The Debye frequency ($\omega_{D}$) of the Wigner lattice can be
estimated within the harmonic approximation as $\omega_{D}$
$\approx$ (e$^2$/$\epsilon_{0}$m$^*$L$^3$)$^{1/2}$ $\sim$ 42
cm$^{-1}$ ($\sim$ 7.8 x 10$^{12}$ s$^{-1}$) which gives the Debye
temperature ($\theta_{D}$) of 60 K with $\epsilon_{0}$ $\sim$ 10,
the static dielectric constant of the underlying CuO$_2$ lattice,
and m$^*$ $\sim$ 60m$_e$. Therefore, we expect the linear
T-dependence down to 0.2$\theta_{D}$ $\sim$ 12 K $^{29}$ since
the main energy dissipation channel for electrons is the
scattering with the acoustic phonons of the Wigner lattice.
Through the oscillator strength calculation, we found only $\sim$
0.42 \% of the doped carriers are contributing to the Drude-like
conductivity at T = 300 K which is consistent with the value
estimated from the plasma frequency. This is 50\% of the
available free carriers at {\it p} = 0.063 doping. At 300 K, the
other 50\% may occupy the single particle excited states of the
Wigner lattice that might contribute to the incoherent background
conductivity. The deviation from the linear T-dependence of the
resistivity in under-doped cuprates is due to additional
scattering with the lattice defects and domain boundaries whose
fluctuation is dynamic as the Wigner lattice formation is not
complete at finite temperatures.

$\sigma_{1}$($\omega$) of $\delta$ = 0.032 sample, which
undergoes the superconducting transition at T$_c$ $\sim$ 26 K, is
displayed in Figure 5a. For $\delta$ = 0.032 sample, there are
two more structures in $\sigma_{1}$($\omega$) in addition to
those found in $\delta$ = 0.024 sample: A sharp peak at $\omega$
$\sim$ 46 cm$^{-1}$ and a broader one peaked at $\omega$ $\sim$ 70
cm$^{-1}$. Since we know that the $\delta$ = 0.032 sample is
right at the fast transition region from a T$_c$ $\sim$ 15K state
to a T$_c$ $\sim$ 30K state. It is plausible that the observed
$\sigma_{1}$($\omega$) of $\delta$ = 0.032 sample comes from two
different contributions; one comes from the T$_c$ = 16 K phase
(LT$_c$) seen in $\delta$ = 0.024 sample and the other from the
T$_c$ = 30 K phase (HT$_c$). Then, the two new features belong
to HT$_c$ phase and have their origin in a different Wigner
lattice ground state. We identify that the $\sim$ 46 cm$^{-1}$
mode corresponds to the collective mode of this new Wigner
lattice, $\omega_{GH}$.

\begin{figure}[tbh]  
\begin{center}
\epsfig{file=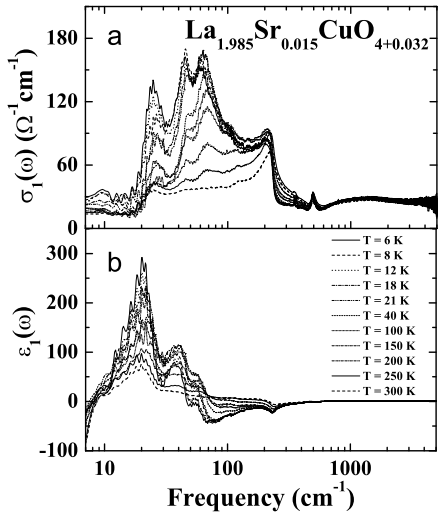,bb=11 11 219 256,width=5.8cm,clip}
\caption{(a) Real part of the conductivity $\sigma_{1}$($\omega$)
and (b) real part of the dielectric function
$\epsilon_{1}$($\omega$) of $\delta$ = 0.032 sample.} 
\end{center} 
\end{figure}

From the oscillator strengths of the Goldstone modes, one may
determine the ratio of the carrier density in LT$_c$ phase (n$_L$)
and HT$_c$ phase (n$_H$), n$_H$/n$_L$. At 30 K, we find
n$_H$/n$_L$ $\sim$ 2, which suggests that the carrier
concentration of HT$_c$ phase is twice of that of the LT$_c$
phase when we assume the same m*. This implies that the HT$_c$
phase contains c(2x2) Wigner lattice, corresponding to {\it p} =
1/8. That is, the degree of commensuration has been decreased by
a factor of two and the corresponding Goldstone mode $\omega_{GH}$
= 2$\omega_{GL}$. We suggest that the physical origin of the
so-called {\it p} = 1/8 anomaly lies in the fact that at
{\it p} = 1/8, the entire holes in the cuprate participate in
forming the HT$_c$ phase Wigner lattice. However, we found n$_H$/n$_L$
$\sim$ 1.2 at T = 300 K and $\sim$ 2.0 at 150 K suggesting that
the HT$_c$ phase grows and builds up the long range order as T is
lowered below 300 K. The $\sim$ 70 cm$^{-1}$ peak in the HT$_c$
phase corresponds to the $\sim$ 90 cm$^{-1}$ peak in the LT$_c$
phase. In this new lattice, the Coulomb band gap has increased
slightly owing to the decrease in the lattice size of the Wigner
lattice.

Based on all the above observations, the normal state
properties of cuprates are characterized as following:

(1) Away from half-filling, the new ground state is the 2D Wigner
lattice order state and there exists a Coulomb band. Only a
small fraction of carriers ($\sim$ 0.42 $\%$ at 300 K and $\sim$
0.97 $\%$ at 150 K) in the Coulomb band are participating in the
coherent charge transport.

(2) Consequently, there exists two different types of gap; one is
the single particle excitation gap at 2$\Delta$ $\sim$ 400 cm$^{-1}$ and
another is the Coulomb band gap on the order of 300 cm$^{-1}$. We
believe this 2$\Delta$ is the gap seen in angle resolved photoemission
spectroscopy (ARPES), tunneling, and heat capacity measurements,
which has been identified as the normal state pseudogap.$^{5}$ In particular, since the best spectral resolution of
ARPES is $\sim$ 20 meV ( $\sim$ 160 cm$^{-1}$),$^{30}$ the
lower energy excitations than 2$\Delta$ have not been resolved in ARPES
measurements. However, the observed states within the gap in
other experiments have their origin in the density of states of
the Coulomb bands.

(3) The Coulomb interaction between free carriers in the Coulomb
band is governed by the dielectric function arising from the
presence of the pinned Goldstone mode through
\begin{equation}
$$\epsilon_{1}(\omega) = 1 +
4\pi n_{W}e^2/m^{*}(\omega^2_{G} -
\omega^{2}\,)\,+\,\epsilon_1\,(\,2\Delta\,\,)$$
\end{equation}
Here, $\omega$$_G$ is the Goldstone mode, n$_W$ is the carrier
density in the Wigner lattice and $\epsilon_1$(2$\Delta$) is the
contribution from the single particle excitation.

Having found that the ground state of cuprates away from
half-filling is the Wigner lattice state and the long range
Coulomb interactions among the excess carriers in the Coulomb
band are massively screened by the presence of the Wigner
lattice, any purely attractive interaction will induce the spin
singlet pair formation which ultimately gives the superconducting
order. The interaction energy between two electrons can be
calculated by considering the following Coulomb interaction
matrix element between two holes, $\langle$\,H$_{12}$\,$\rangle$ which takes the form 
\begin{equation}
$$\langle H_{12} \rangle =
4\pi e^{2}\,/\,q^{2}\epsilon_1\,(q,\omega)$$ 
\end{equation}
Since the holes in the Coulomb band also oscillates in phase with
the Wigner lattice at $\omega$ = $\omega$$_G$ and
1{\bf /}$\epsilon_1$\,{\bf (}\,q\,$\rightarrow$\,0, $\omega${\bf )}\,$\approx$\,1\,-\,$\omega$$^2_0$\,{\bf /}{\bf (}\,$\omega$$^2_G$\,
-\,$\omega$$^2$\,+\,$\omega$$^2_0$\,{\bf )} in the long
wavelength limit with $\omega$$^2_0$\, $\equiv$
4$\pi$n$_W$e$^2$/m$^*$, one immediately finds the range of
$\omega$ where V = $\langle$H$_{12}$$\rangle$ is
attractive as $\omega$$_G$ $<$ $\omega$ $<$
$\sqrt{\omega^{2}_{0} + \omega^{2}_{G}}$. Here we have ignored the
contribution from the single particle excitation and the damping
of $\omega$$_G$. Notice that this interaction is not retarded.
Therefore, the free carriers in the Coulomb band already form
pairs once the long range order of the Wigner lattice develops
for temperatures below T $\sim$ $\omega$$_0$ $\sim$ 215 K. Only
when the photon energy greater than $\omega$$_0$ (or T
$>$ $\sim$ 215 K), these preformed pairs dissociates via the Coulomb
repulsion. In fact, vortex-like excitations in
La$_{2-x}$Sr$_{x}$CuO$_{4}$ systems at T as high as 150 K have been
observed in the measurements of the Nernst effect. $^{31}$

How does the superconducting order set in at T = T$_c$$?$ There
are evidences supporting that the phase stiffness of the
superconducting order parameter $\Psi$ $\sim$
$\Delta$e$^{i \phi}$ plays a crucial role in high T$_c$ cuprates.$^{32}$ Even though the pairing of the holes is achieved, the
superconducting order will be obtained only after the long-range
order in phase has been established. In conventional BCS
superconductors, because of the phase stiffness energy is much
greater than the pair binding energy, the transition to the
normal state is achieved by breaking the pair at T $\geq$ T$_c$
and the superconducting transition and pairing take place at the
same temperature.

In the original BCS theory, the derivation of the Meissner
effect from the reduced BCS Hamiltonian is not gauge invariant.
From consideration of gauge invariance, Bogoliubov$^{33}$
pointed out that there exist collective excited states of
quasi-particle pairs, which can be excited only by the
longitudinal current associated with the longitudinal component
of the vector potential. Therefore, once the superconducting
order has been established in ordinary BCS superconductors, the
phase collective mode is present due to the broken gauge symmetry
and the presence of such collective states is essential for the
gauge invariance in the long wavelength limit (q $\rightarrow$ 0).

Nambu$^{34}$ found that the charge density correlation in
the ground state is given by X(q,\,$\omega$) $\approx$
n$_F$q$^2$/m$_e$\,($\omega^2$ -
n$_F$q$^2$/\,N(0)\,) where n$_F$ = free electron density and
N(0) = density of states. In the presence of the Coulomb
interaction between electrons, the dispersion relation for the
collective excitations can be determined from the secular
equation, 1 - V(q)X(q,\,$\omega$) = 0 with bare Coulomb interaction
V(q) = 4$\pi$e$^2$/q$^2$. This secular equation gives
$\omega_{\phi}$$^2$ $\sim$ 4$\pi$n$_F$e$^2$/m$_e$ in q
$\rightarrow$ 0 limit which is the plasma frequency of the
electron gas. This collective mode is known as the
Bogoliubov-Anderson mode.$^{35}$ This mode shifts to the
plasma frequency makes the observation of the phase collective
mode impossible. However, for the case of the Wigner lattice
ground state, we propose that one must consider the screened
Coulomb interaction in the Wigner lattice ground state as
\begin{equation}
$$V(q,\omega) = 4\pi e^2/\,q^2\epsilon_1(q,\omega)$$
\end{equation}
instead. This modification leads to the phase collective mode
solution $\omega$$_{\phi}$$^2$ $\approx$
$\omega$$_G$$^2$\,/\,(1\,+\,$\gamma$) in q $\rightarrow$ 0 limit
with $\gamma$ $\equiv$ n$_W$m$_e$\,/\,n$_F$m$^*$ for the
electrons pairs in the Wigner lattice with $\epsilon_{1}$
($\omega$) $\approx$ 1 +
4$\pi$n$_{W}$e$^2$\,/\,m$^*$($\omega_{G}^{2}$ - $\omega^{2}$).
This implies that the phase collective mode $\omega$$_{\phi}$
occurs when the pairs of electrons in 2D Wigner lattice undergo
the Kosterlitz-Thouless transition at T = T$_c$.

The temperature dependence of the oscillator strength of the
Goldstone modes and the Coulomb bands of $\delta$ = 0.024 and
$\delta$ = 0.032 samples, plotted in Figure 6 and Figure 7
respectively, are consistent with this picture. Two main
remarkable changes are clearly seen; one is a relatively abrupt
increase in the oscillator strength at $\sim$ 200 K, common to all
the peaks and the other is a sharp jump in the strength of the
Goldstone modes, $\omega_{GL}$ and $\omega_{GH}$, at their
respective superconducting T$_c$, one at 16 K and the other at 30
K. The increase in the oscillator strength of the collective
modes at 200 K, indicates that the long-range order of the Wigner
lattice starts to develop as evidenced by the reduction in their
dynamic mass. At the same time, the strength increase in the
Coulomb band suggests that more carriers occupy the Coulomb band
following the development of the long-range order of the Wigner
lattice. The corresponding free-carrier oscillator strength
changes from $\sim$ 0.42 \% at 300 K to $\sim$ 0.97 \% at 150 K
and to $\sim$ 1.18 \% at 30 K in $\delta$ = 0.024 sample
indicates that all the available free carriers are now occupying
the Coulomb band and participating in the charge transport at T
below 200 K. However, we do not observe the change in the free
carrier plasma frequency because of the increasing static
dielectric constant with decreasing T. In $\delta$ = 0.032
sample, the free carrier strength is $\sim$ 0.44 \% of the total
carrier at 300 K and $\sim$ 0.71 \% at 30 K. At the same time, as
demonstrated above, at the crossover temperature (T$_0$ $\sim$ 200
K) of the long-range order of the Wigner lattice, the free
carriers in the Coulomb band start to form pairs. Subtle changes
in the d$\rho$/dT vs. T curves of both $\delta$ = 0.024 and
$\delta$ = 0.032 samples at $\sim$ 200K has already been pointed out
(see Figure 2). The resistivity decreases faster and faster upon
cooling below 200 K. Therefore, we have normal state pairs of
holes formed at 200 K in our cuprate system. It is interesting to
note that 200 K is exactly the temperature when mobile excess
oxygen is considered to be frozen in.

\begin{figure}[tbh]  
\begin{center}
\epsfig{file=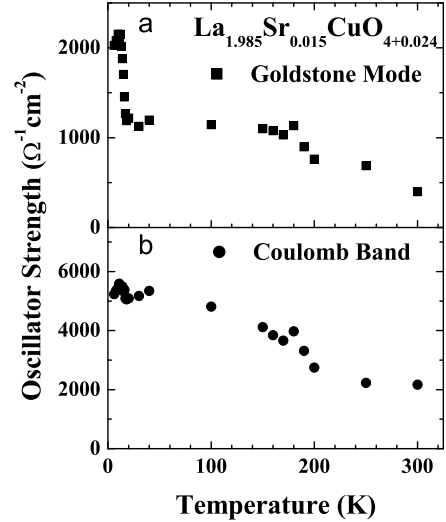,bb=11 11 219 263,width=5.8cm,clip}
\caption{Oscillator strength of (a) the Goldstone mode and (b)
the Coulomb band of $\delta$ = 0.024 sample versus temperature
plot.} 
\end{center} 
\end{figure}

\begin{figure}[tbh]  
\begin{center}
\epsfig{file=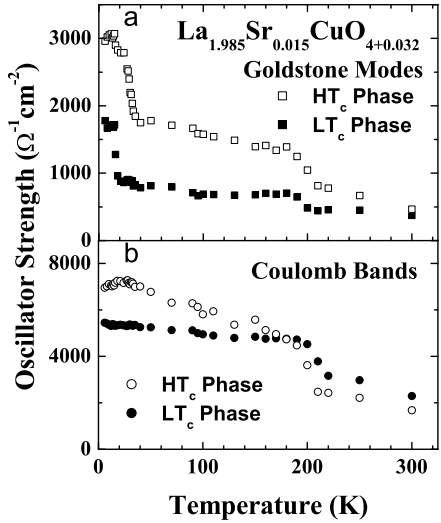,bb=11 11 220 262,width=5.8cm,clip}
\caption{Oscillator strengths of the two Goldstone modes, one for
T$_c$ = 16 K (LT$_c$ phase) and the other for T$_c$ = 30 K (HT$_c$ phase), and the corresponding oscillator strength of the Coulomb bands of $\delta$ = 0.032 sample. Each oscillator strength was calculated from the real part of the conductivity by fitting with a symmetric Gaussian function. (See the text for details.)} 
\end{center} 
\end{figure}

We propose that the abrupt increase of the oscillator strength
of the Goldstone mode at T$_c$ is resulted from the development of
the phase collective mode at T = T$_c$ under $\gamma$ $\ll$ 1 condition
(m$^*$ $\gg$ m$_e$ and n$_W$ $>$ n$_F$), which gives $\omega_{\phi}$
$\sim$ $\omega_G$. The abrupt increase of the oscillator strength
of $\omega_{GL}$ mode at 16 K and $\omega_{GH}$ mode at 30 K
confirms the development of the phase collective mode of the
superconducting order as the phase coherence develops at T$_c$.
Below T$_c$, the phase of the superconducting order parameter
propagates as a collective mode. From the data, T$_c$ $\sim$ 16 K
is measured with $\hbar\omega_{GL}$ $\approx$ 23 cm$^{-1}$ and
T$_c$ $\sim$ 30 K with $\hbar\omega_{GH}$ $\approx$ 46
cm$^{-1}$. Therefore the superconducting T$_c$ can be predicted by measuring the
Goldstone mode frequency via k$_B$T$_c$ $\approx$
0.5$\hbar\omega_{G}$. In fact, the phase collective mode and
the Goldtone mode (or the hint of their presence) have been
observed in a number of cuprate systems.$^{36-47}$
Specifically, we believe that the mode observed at $\sim$ 80
cm$^{-1}$ in the underdoped Bi$_2$Sr$_2$CaCu$_2$O$_{8+\delta}$
(Bi2212) (T$_c$ = 67 K) by Timusk et al.$^{43}$ is the phase
collective mode associated with the pinned Goldstone mode at $\sim$ 93
cm$^{-1}$ with $\gamma$ $\sim$ 2.8. Also, in their far-infrared study of
Tl$_2$Ba$_2$CuO$_{6+\delta}$, Timusk et al.$^{44}$ reported an
observation of a peak at 70 cm$^{-1}$ below T$_c$ = 88 K which
accompanies a satellite peak at $\sim$ 120 cm$^{-1}$. This 70
cm$^{-1}$ mode is the phase collective mode with $\gamma$ $\sim$
2.0. The large $\gamma$ implies the reduction in the dynamic
mass of the Wigner lattice compensated by the increase in n$_F$,
free carrier density. Since the oscillator strength of the single
particle excitation may be approximated as S$_{2\Delta}$ $\sim$
$\pi$ne$^2$(1-n$_{W}$m$_e$/nm$^*$)/2m$_e$ with n = n$_W$ + n$_F$,
the reduction in the dynamic mass of the lattice diminishes the
oscillator strength as actually seen in their in-plane
conductivities.

In summary, focusing on the special hole concentration
{\it p}$_c$ = 1/16, we have studied the charge dynamics of Sr-
and O- co-doped La$_{1.985}$Sr$_{0.015}$CuO$_{4+\delta}$ with
$\delta$ = 0.024 ( {\it p} = 0.063 per Cu) and $\delta$ = 0.032
( {\it p} = 0.07). We observed the Goldtone mode of p(4x4)
Wigner lattice corresponding to {\it p}$_c$ = 1/16 for {\it p}
= 0.063 sample and the Goldstone mode associated with c(2x2)
Wigner lattice ({\it p}$_c$ = 1/8) in addition to the
contributions of the p(4x4) lattice for {\it p} = 0.07 sample.
We found that the presence of 2D Wigner lattice and the pinned
Goldstone mode is essential for the cuprate physics and
superconductivity. We propose that all the high T$_c$ physics are
based on the existence of these peculiar 2D electron lattices which are realized by carrier doping into an antiferromagnetic background.

{\bf Acknowledgment} We would like to thank Zugang Li, Zheng Wu and
Young Seok Song for their various technical assistances as well
as sample preparations and characterizations during the entire
course of this study. We thank John Markus for his genuine
interest and technical helps to expedite our experiment. We are
also indebted to Bob Hansen at R.G. Hansen and Associates for
kindly lending us their liquid helium transfer line during the
critical moment of this experiment. PHH is supported by the State
of Texas through The Texas Center for Superconductivity at the
University of Houston.

\end{document}